\documentclass[11pt,twoside]{article}

\usepackage{asp2006}
\usepackage{epsf}
\usepackage{psfig}
\usepackage{lscape}
\usepackage{graphicx}

\begin{document}
\title{GOLF: a new proxy for solar magnetism}   
\author{S. Mathur}   
\affil{Indian Institute of Astrophysics, Koramangala, Bangalore 560034, India}    
\affil{Instituto de Astrof\'isica de Canarias, 38205, La Laguna, Tenerife, Spain}
\author{S.~J. Jim\'enez-Reyes}
\affil{Instituto de Astrof\'isica de Canarias, 38205, La Laguna, Tenerife, Spain}
\author{R.~A. Garc\'\i a}   
\affil{Laboratoire AIM, CEA/DSM-CNRS-Universit\'e Paris Diderot; CEA, IRFU, SAp, Centre de Saclay, F-91191, Gif-sur-Yvette, France}    

\begin{abstract} 
Solar magnetism is measured with different indexes: for instance the MPSI and the MWSI, number of sunspots, radio flux at 10.7 cm, Ca II K, Mg II K, EUV, He I or L$_\alpha$. Bachmann \& White (1994) had compared these indicators of the solar activity showing a hysteresis of the solar cycle variations and a time lag between these indices not related to instrumental effects. Later on, Ozg\"u\c c \& Ata\c c (2001) extended this study of hysteresis phenomenon between Flare index and other solar indices (mean magnetic field, coronal index). In its original working configuration, GOLF/SoHO was able to measure during 26 days the solar mean magnetic field (Garc\'ia et al. 1999). We check here if the velocity data could be used as another solar magnetism proxy with the advantage of having a duty cycle $\ge$95\% during the last 12 years. We will compare the GOLF data with some of the above-mentioned solar activity indexes.
\end{abstract}



\section{Introduction}
GOLF\footnote{Gabriel et al. 1995} is a resonant scattering spectrophotometer dedicated to the study of low-degree acoustic (e.g. Garc\'\i a et al. 2001) and gravity modes (e.g. Mathur et al. 2007; Garc\'\i a et al 2008). With a sampling time of 10s, it measures the Doppler velocity on the doublet sodium line, NaD1 and NaD2, by integrating two points of $\sim$ 25 $m\AA$ width on the same wing. During the last 12 years, measurements have been alternated between the Blue Wing (BW), the Red one (RW) and back to the Blue one (Garc\'ia et al. 2005). Each observed wing corresponds to a different height in the solar atmosphere (Jim\'enez-Reyes et al. 2003, 2007). In the presence of a magnetic field the solar Na absorption lines are broaden (Robillot, Bocchia and Denis 1993) and, therefore, the Doppler velocity observed by GOLF will be sensitive to the magnetism at the surface of the Sun.

\section{Time and spectral analyses and comparison with solar indexes}

\begin{figure}[htb*]
	\centering
	\begin{tabular}{cc}
		\includegraphics[width=6.5cm]{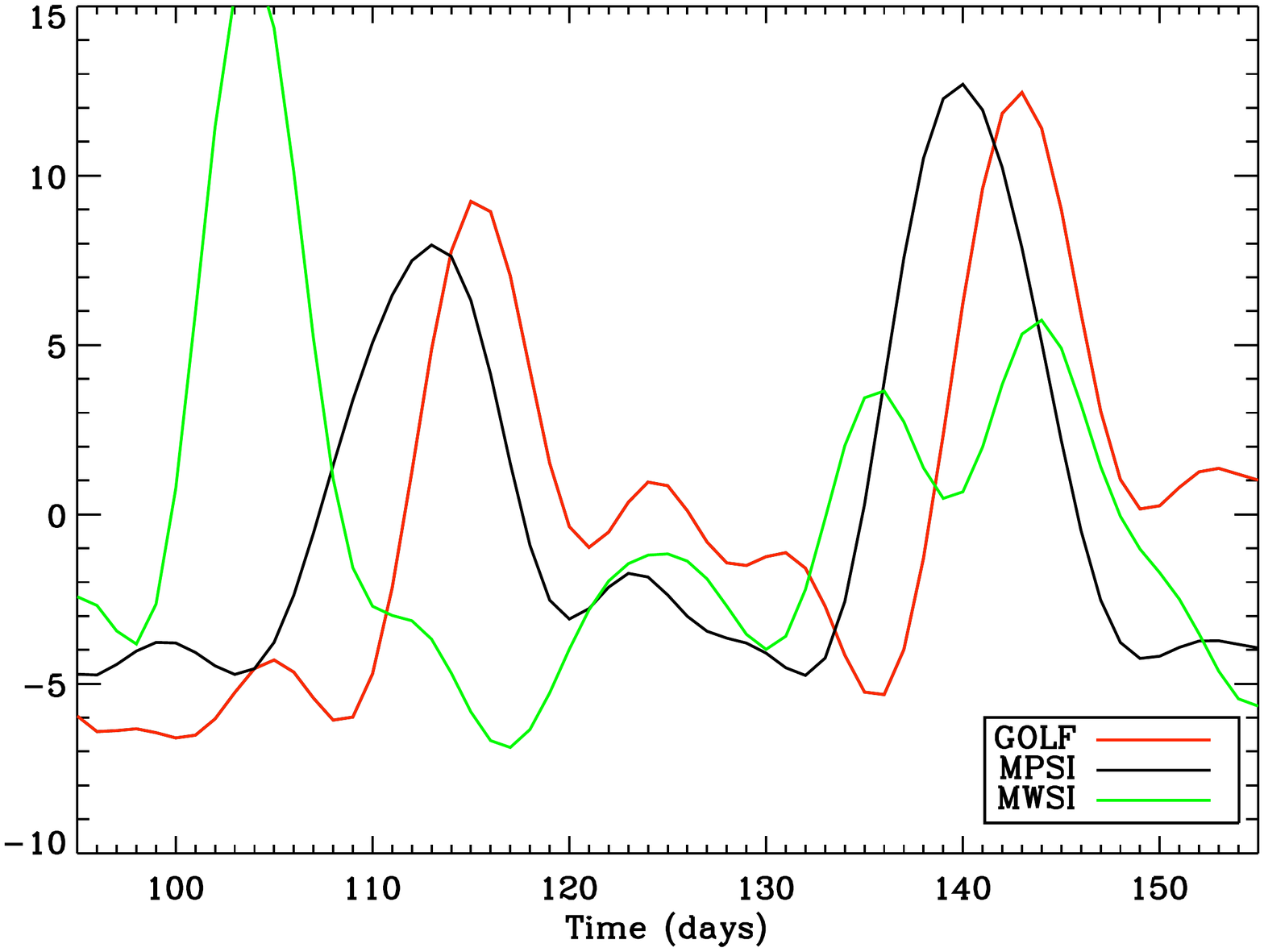} &
		\includegraphics[width=6.5cm] {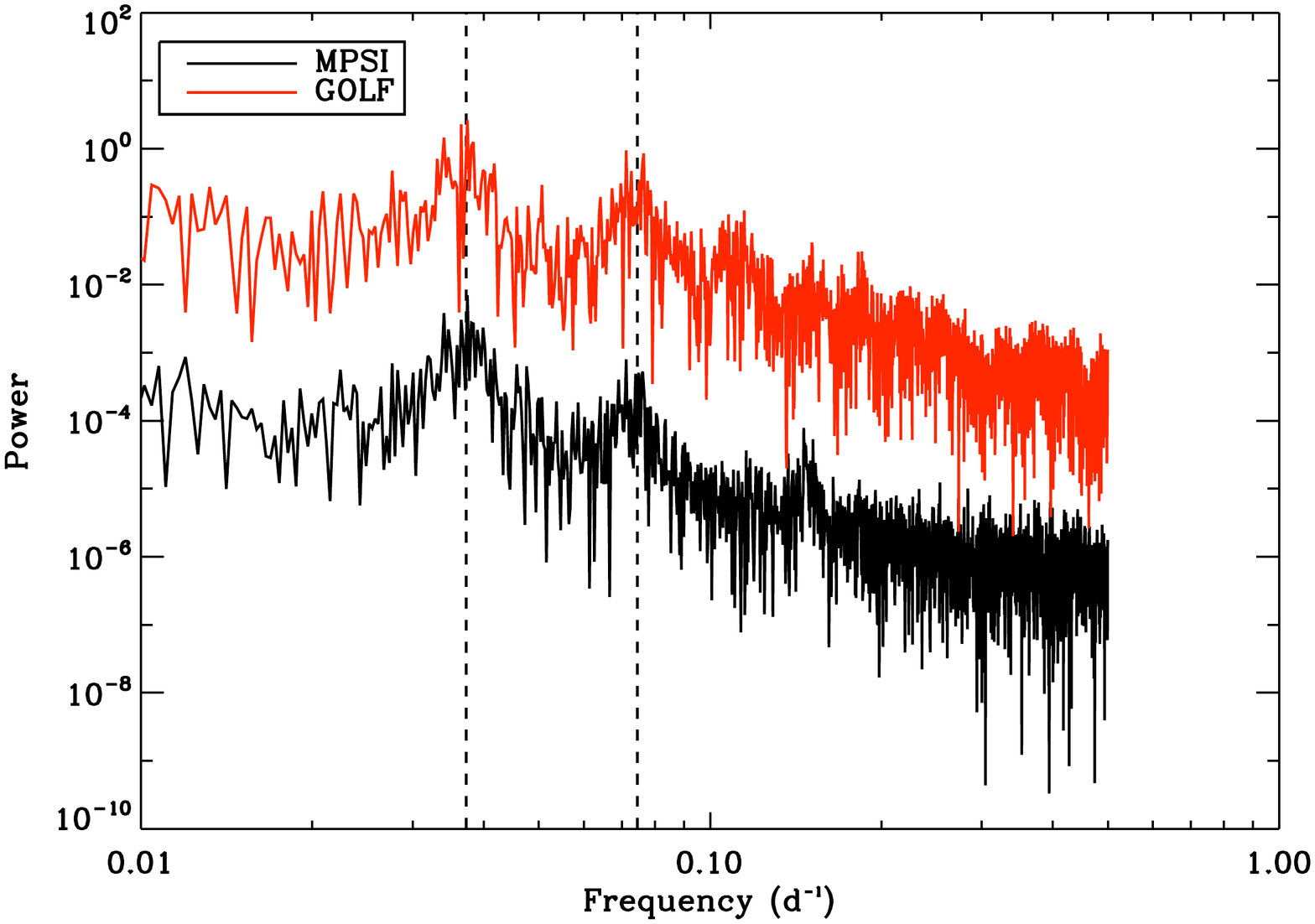}

	\end{tabular}
	\caption{On the left: zoom of the time series of GOLF, MPSI and  MWSI. On the right: Power Spectrum Density at low frequency of GOLF and MPSI.
}
	\label{temp_spec}
\end{figure}

Two of the usual solar activity indexes are the Magnetic Plage Solar Index (MPSI) and the Mont Wilson Sunspots Index (MWSI). The MPSI is the sum of those  pixels where the absolute value of the magnetic field is between 10 and 100 Gauss. Whereas the MWSI is calculated for pixels having absolute values greater than 100 Gauss. \\
GOLF and MPSI see the same magnetic structures but shifted in time, whereas the magnetic structures in the MWSI are completely different (Fig.~\ref{temp_spec} (left)). Indeed GOLF integrates the full solar disk and thus takes a mean value which is mainly the weak magnetic field as sunspots only represent a few percent of the solar surface. Thus GOLF is more sensitive to small magnetic structures. Meanwhile, MPSI and GOLF spectra present two peaks corresponding to the ~26 days periodicity (rotation period) and its first harmonic (Fig.~\ref{temp_spec} (right)).

\begin{figure}[!h]
\centering
\includegraphics[width=7cm]{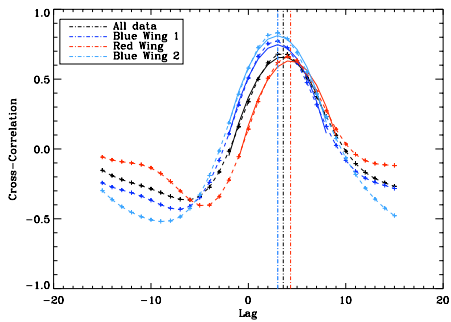}
\caption{Cross-Correlation between GOLF and the MPSI for different configuration of observation of GOLF as a function of the time lag in days.}
\label{cross_cor}
\end{figure}

We deepen our study by using two more indexes: sunspot numbers (SS) and the radio flux at 10.7 cm (F10.7). We calculate the correlation rates and time lags between {them and GOLF}. Table~\ref{tab} shows that the observation of activity is made at different moments depending on the index studied. GOLF data present a delay from 3 to 5 days compared to the other indexes. Moreover, the correlation rate is higher with MPSI than with MWSI or SS as GOLF integrates solar light and is sensitive to weak magnetic field (see Section~2).\\
For the different periods of time in GOLF observations, we calculate the time lags and correlation coefficients between GOLF and MPSI (Fig. \ref{cross_cor}). Using 12 years of data, the correlation has a maximum of 69\% for a time lag of 3.57 days, meaning that magnetic structures are seen 3.57 days after by the GOLF instrument. This is probably due to a lack of sensitivity towards the solar limb (Garc\'\i a et al. 1998). The decrease of the correlation  in the RW is due to the lower sensitivity of this operating mode to the broaden of the line in presence of magnetic field.

\begin{table}[!h] 
\caption{Time lag and correlation coefficient between GOLF and five solar activity indexes} 
\smallskip 
\begin{center} 
{\small 
\begin{tabular}{ccccc} 
\tableline 
\noalign{\smallskip}
Index & Time lag (days) & Correlation rate\\
\noalign{\smallskip} 
\tableline 
\noalign{\smallskip}
MPSI & 3.57 & 69\%\\
MWSI & 3.62 & 48\%\\
MPSI+MWSI & 3.58 & 68\%\\
SS & 3.45 & 51\%\\
F10.7 & 4.95 & 53\%\\
\noalign{\smallskip} 
\tableline 
\end{tabular} 
}
\end{center} 
\label{tab}
\end{table}

\section{Using the wavelet technique}
\begin{figure}[htbp]
\begin{center}
\includegraphics[width=12cm, height=2cm]{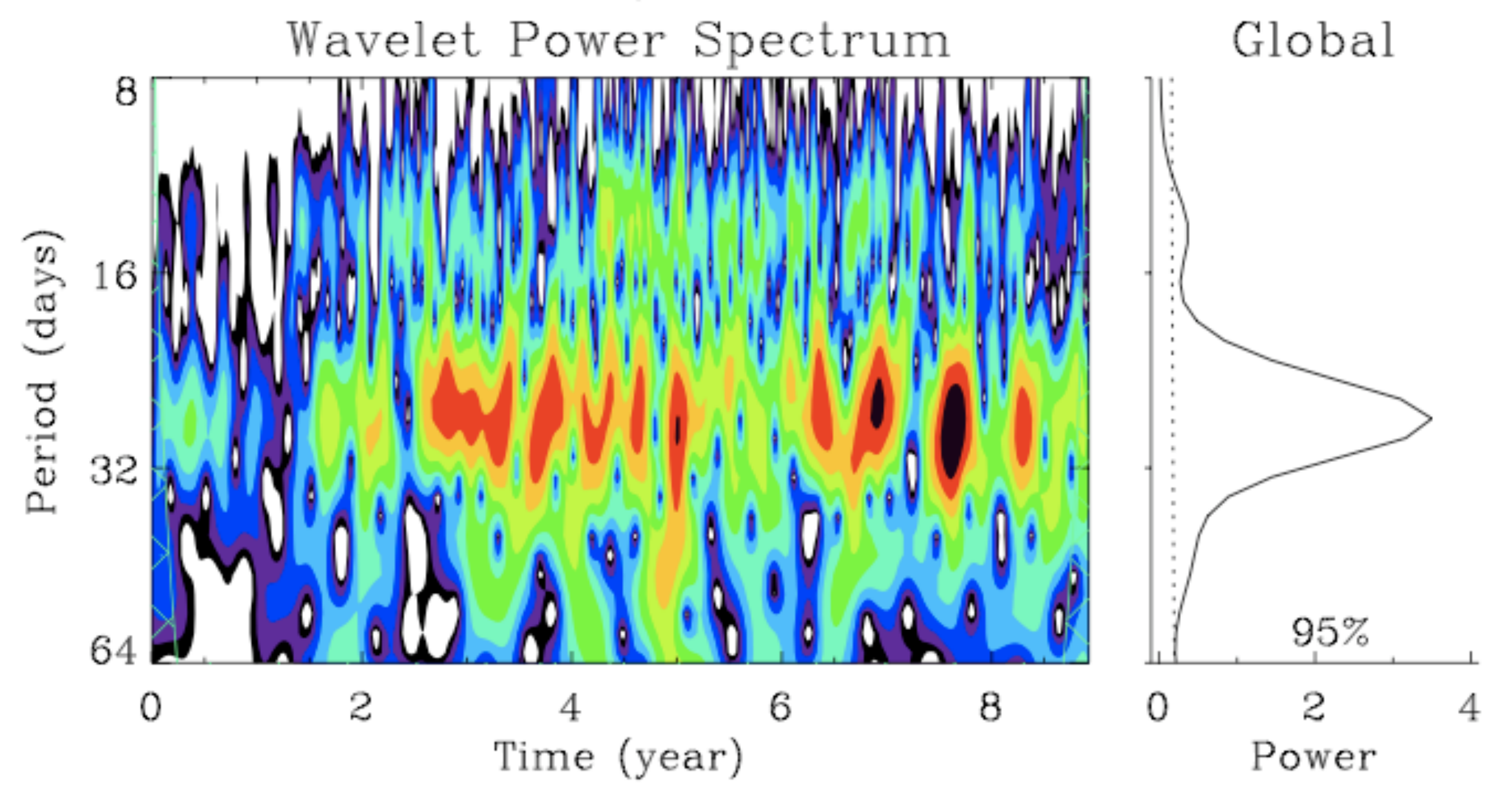}
\includegraphics[width=12cm, height=2cm]{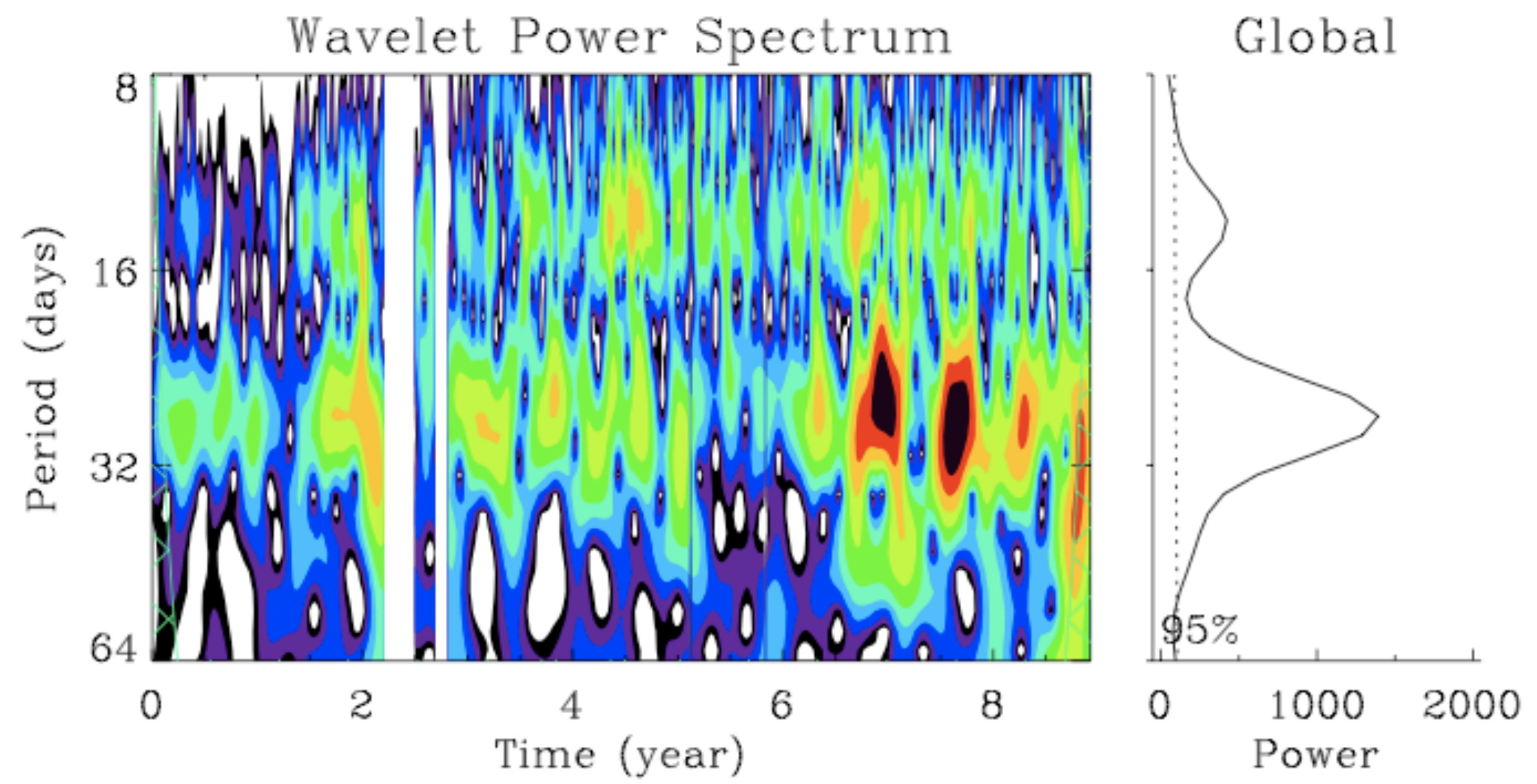}
\includegraphics[width=8.5cm, height=2.3cm]{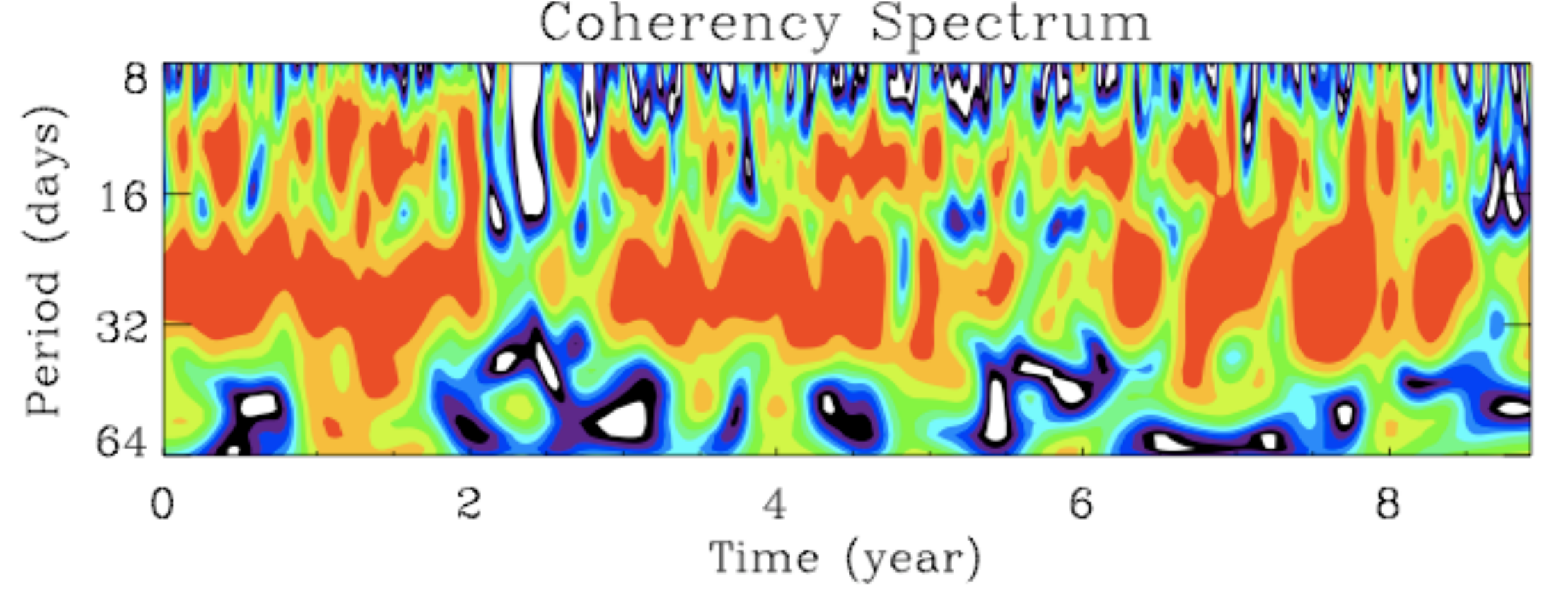}
\caption{Wavelet Power Spectrum as a function of time for the MPSI (top panel) and GOLF data (middle panel). Wavelet power spectrum coherency between GOLF and MPSI data (bottom panel).}
\label{wave}
\end{center}
\end{figure}

Applying the wavelet\footnote{Wavelet software was provided by C. Torrence and G. Compo and is available at URL: http://paos.colorado.edu/research/wavelets/} technique to the data, we obtained the Wavelet Power Spectrum (Fig.~\ref{wave} top and middle) as a function of the period of the wavelet and  of time. Regions where the power is more significant (in red and black colours) appear in both spectra and almost at the same moment. The same behaviour is observed with the other solar activity indicators. Fig.~\ref{wave} (bottom) shows the coherency between the wavelet power spectra of MPSI and GOLF data shifted by 3.57 days. The maximum coherency is observed around a period of 26 days.

\section{Conclusions}

A 26-days periodicity appears in GOLF data (peak at 26.8 days and first harmonic at 13.4 days in the Power Spectrum). This signal had been seen by Claverie et al. (1982, 1983) and correctly interpreted later by Duvall et al. (1983) and Edmund \& Gough (1983). Larger periodicities cannot be found as a low-frequency filter is applied to GOLF data. GOLF is sensitive to weak magnetic field as it integrates the full solar disk. However, GOLF observes solar magnetism a few days after other classical activity indexes but with a very high duty cycle ($\ge$95\% in 12 years) compared to the MPSI ($\sim$75\% during the same period) and with a very fast cadence (dt=10s).\\
The wavelet analysis of the integrated velocity could be a very good tool in asteroseismology to obtain an indication of the full projected orbital velocity and not half of the rotation period as it is the case with the power spectral density.

\acknowledgements 
This work has been partially funded by the Spanish grant PENAyA2007-62650 and the CNES/GOLF grant at the SAp-CEA/Sacaly. SOHO is a cooperation between ESA and NASA.


\end{document}